\documentclass[aps,preprint,groupedaddress,showpacs]{revtex4}
\newcommand{\be}{\begin{equation}}
\newcommand{\ee}{\end{equation}}
\newcommand{\bea}{\vspace{0.25cm}\begin{eqnarray}}
\newcommand{\eea}{\end{eqnarray}}

\def\PLA{{Phys. Lett.}  A }
\def\PRL{{Phys. Rev. Lett.} }

\begin{document}

\title{ A first experimental test of de Broglie-Bohm theory against standard 
quantum mechanics }

\author{G. Brida} 
\author{E. Cagliero}
\author{G. Falzetta}
\author{M. Genovese}
\email{genovese@ien.it}
\homepage{http://www.ien.it/~genovese/marco.html}
\author{M. Gramegna }
\author{C. Novero }
\thanks{ Dedicated to the memory of our beloved friend and collaborator Carlo 
Novero.}

\affiliation{ Istituto Elettrotecnico Nazionale  Galileo Ferraris, Strada delle 
Cacce 91, 10135 Torino, Italy }

\date{\today}

\begin{abstract}
 Quantum Mechanics is a pillar of modern physics, confirmed by a huge amount of 
experiments. Nevertheless, it presents many unintuitive properties, strongly 
differing from classical mechanics due to its intrinsic non-epistemic 
probabilistic nature. Many attempts have been devoted to build a deterministic 
theory reproducing all the results   of Standard Quantum Mechanics (SQM), but 
where probabilities are epistemic, namely due to our ignorance of some hidden 
variables. These theories can be local or non-local. The formers are 
substantially excluded by Bell inequalities experiments \cite{1,3}. The latter 
include the  de Broglie - Bohm (dBB) one \cite{1,8}, the most successful attempt 
in this sense. DBB theory is built in order to reproduce almost all  SQM 
predictions. However, it has recently been shown \cite{4,5,6} that different 
coincidence patterns are predicted by SQM and dBB when a double slit experiment 
is realised under specific conditions, in particular no coincidence is expected 
when the two photodetectors are in the same semiplane respect to the median 
symmetry axis of the double slit. In this letter we present the first 
realisation of such a double slit experiment with correlated photons produced in 
type I parametric down conversion. We observe a perfect agreement with SQM 
prediction and a coincidence peak almost 8 standard deviations above zero, when the 
two photodetectors are well inside the same semiplane: Thus our results confirm 
Standard Quantum Mechanics contradicting dBB predictions. 

\end{abstract}


\maketitle

\section{Introduction}

Quantum Mechanics represents a pillar of modern physics: an impressive amount of 
experiments have confirmed this theory and many technological applications are 
based on it. 

According to its  standard interpretation, Quantum Mechanics nature is 
intrinsically probabilistic, permitting only predictions about probabilities of 
the occurrence of an event \cite{1}. The state of a system is described by a 
wave function, whose modulus squared gives the probability density distribution 
of the system. Of course, in this picture, physical systems do not follow 
trajectories in their motion because of their position is not perfectly defined. 
The non-epistemic nature of probabilities in SQM leads to many unintuitive 
properties of quantum systems and, more badly than that, the transition between 
the microscopic quantum world and macroscopic deterministic classical world 
remains unsolved:  None of the many attempts to give a solution of macro-
objectivation \cite{1} has effectively reached a universal consent in physicists 
community.

Starting from an Einstein's \cite{2} work, many attempts have been devoted to 
build a deterministic theory reproducing all the results of SQM, in which 
probabilities become epistemic, namely are due to our ignorance of some hidden 
variables, whose knowledge would in fact make the evolution of the system 
perfectly determined.

A fundamental paper of Bell \cite{9} showed that local Hidden Variable Theories 
(HVT) cannot reproduce all the results of SQM. Since then a lot of experiments 
have been addressed to test local HVT against SQM. All of them \cite{1,3} 
confirmed SQM, even though a conclusive experiment is still missing due to the 
low detection efficiency, which leads to the need of an additional hypothesis 
stating that the final measured sample is a faithful representation of the 
initial one (detection loophole) \cite{santos}.

However, Bell theorem does not apply to non-local hidden variable theories,   
including the de Broglie-Bohm one \cite{1,7,8}, the most serious and successful 
attempt to propose an alternative to SQM. In this theory the hidden variable is 
the position of the particle, which follows a perfectly defined trajectory in 
its motion. The evolution of the system is given by classical equations of 
motion, but an additional potential  must be included. This "quantum" potential 
is related to the wave function of the system and thus it is non-local. The 
inclusion of this term, together with an initial distribution of particles 
positions given by the quantum probability density,  successfully allows the 
reproduction of almost all the predictions of quantum mechanics.

Nevertheless, in a  series of recent papers \cite{4,5,6} it has been shown that 
under particular conditions (namely when the system is not ergodic \cite{14}) 
different experimental results are predicted by dBB and SQM, allowing a test of 
the two theories. In particular, in dBB  one expects that \cite{4,5,6}, due to 
the non-crossing theorem, in a double slit experiment, where two identical 
particles cross simultaneously each a precise slit, no coincidence can be 
measured in the same semiplane respect to the median symmetry axis of  the two 
slits.

\section{Description of the experiment}

In order to test this prediction  we have realised a novel experimental 
configuration in which two identical photons reach a double slit at the same 
time, each one crossing a different slit, according to the configuration 
proposed  in ref. \cite{4,5,6}. However, it must be pointed out that only in 
some versions of dBB theory the photon is described as a particle 
\cite{4,5,7,8,10}: therefore our test concerns these versions of the theory, 
which have the fundamental advantage of describing bosons and fermions on the 
same ground. If the experiment will be repeated with fermions a conclusive test 
of dBB will be obtained.  In the last years some experiments \cite{11,12,13}  
were performed,  in which PDC light was crossing a double slit, but in none of 
them a median symmetry axis was well defined and a configuration suitable for 
the present test was realized. 

More in details (see fig.1)  a 351 nm pump laser of 0.4 W power is directed into 
a lithium iodate crystal, where correlated pairs of photons are generated by 
type I  parametric down conversion (PDC) \cite{3}. The two photons are emitted 
at the same time (within femtoseconds, whilst correlation time is some orders of magnitude larger) on a well defined direction for a 
specific frequency. By means of an optical condenser the produced photons, 
within two correlated directions corresponding both to 702 nm emission (the degenerate emission for a 351 nm pump laser), are sent 
on a double slit (obtained by a metal deposition on a thin glass by a 
photolithographic process) placed just before the focus of the lens system. The 
two slits are separated by  100  $\mu$m  and have a width of 10 $\mu$m. They lay 
in a plane orthogonal to the incident laser beam and are orthogonal to the table 
plane.  Two EG$\&$G single photon detectors are placed at a 1.21 and a 1.5 m 
distance after the slits. They are preceded by an interferential filter at 702 
nm of 4 nm full width at half height and by a lens of 6 mm diameter and 25.4 mm focal length. 
The output signals from the detectors are routed  to a two channel counter, in 
order to have the number of events on a single channel, and to a  Time to 
Amplitude Converter (TAC) circuit, followed by a single channel analyser, for 
selecting and counting the coincidence events.  

With this experimental set-up, we obtained that a  clear signal appeared on the 
multi-channel analyser when the detectors were placed in the positions where the 
maximum of coincidences was expected, namely at  2 degrees, in two different 
semiplanes,  respect to the median symmetry axis of the double slit (which is 
parallel to the incident pump laser). In order to check that the two degenerate 
photons crossed two different  slits, we have alternatively closed one of the 
slits leaving the other opened: correspondingly, the coincidence peak 
disappeared and the signal on the related detector dropped to background level, 
confirming the correct position of the double slit. Then we moved the first 
detector scanning the diffraction pattern, leaving the second fixed at 5.5 cm 
from the symmetry axis. We found that the coincidences pattern perfectly 
followed quantum mechanics predictions, see fig.2. The last ones are given by:
\bea
& C(\theta_1,\theta_2) = &  g(\theta _1, \theta _i^A )^2 g(\theta _2, \theta 
_i^B)^2 + g(\theta _2, \theta _i^A )^2 g(\theta _1, \theta _i^B)^2 + \cr
& & 2 g(\theta _1, \theta _i^A ) g(\theta _2, \theta _i^B) g(\theta _2, 
\theta _i^A ) g(\theta _1, \theta _i^B) cos [ k s (sin \theta _1 - sin \theta _2)]
\eea
where 
\be 
g( \theta , \theta _i^l)  = { sin ( k w /2 ( sin (\theta) - sin (\theta _i^l)) 
\over 
k w /2 ( sin (\theta) - sin (\theta _i^l))}
\ee
takes into account diffraction. k is the wave vector, s the slits separation, w 
the slit width, $\theta _{1,2}$ is the diffraction angle of the photon observed 
by detector 1 or 2, $\theta_i^l$ the incidence angle of the photon on the slit $l$ (A or B).

Because of finite width of the slit, a partial penetration of trajectories in 
the opposite semiplane is possible, however this penetration is of the order of 
the slit width \cite{4} (10  $\mu$m) and thus negligible. 

The fundamental result of our experiment is that  a  coincidence peak is clearly 
observed (see fig.3) also when the first detector is placed inside enough the same semiplane of the second one. We consider coincidences acquisitions with a temporal window of 2.6 ns, the background is evaluated shifting the delay between start and stop of TAC of 16 ns and acquiring data for the same time of the undelayed acquisition. When the edge of the lens of the first detector is placed -1.4 cm (the center  -1.7 cm) after  the median symmetry axis  of the two slits (the minus means on the left of symmetry axis looking toward to the crystal), whilst the second detector is kept at -5.5 cm, with 35 acquisitions of 30' each we obtained 78 $\pm$ 10 coincidences per 30 minutes after background subtraction, whilst in this situation the dBB \cite{4,5,6} prediction for coincidences is strictly zero. Furthermore, even when the two detectors were placed in the same semiplane the first at -4.44 and the second at -11.7 cm from the symmetry axis, in correspondence of the second diffraction peak, a clear coincidence signal was still observed (albeit less evident than in the former case): We obtained in fact, after background subtraction, an average of 41 $\pm$ 14 coincidences per hour with  17 acquisitions of one hour (and a clear peak appeared on the multichannel analyser). 

%In conclusion, as a final check, we have also measured the coincidences when %the detectors are placed in relative positions such that SQM predicts zero %coincidences as well. Also in this case, i.e. positioning the first detector at %-4.27 cm  and the other one at -13.66 cm respect to the symmetry axis, our %result,  coincidences per hour, was again in perfect agreement with the SQM %prediction.

 Therefore we can conclude that our results are in perfect agreement with SQM predictions, but contradict of almost 8 standard deviations the predictions of dBB theory presented in ref. \cite{4,5,6}.

\section{Conclusions}

In conclusion we have realised the first experimental test of de Broglie-Bohm 
theory against standard Quantum Mechanics following the proposal of ref. 
\cite{4,5,6}.  DBB predicts that in a double slit experiment, in which two 
identical particles simultaneously cross each a precise slit, their trajectories 
remain in the semiplane of the crossed slit and thus no coincidence can be 
measured when two detectors are placed on the same side respect to the median 
symmetry axis. This result is at variance with SQM prediction. In our experiment 
we have clearly observed  a coincidence peak when the detectors are placed in 
the same semiplane, confirming SQM prediction against dBB one.

% figures should be put into the text as floats.
% Use the graphics or graphicx packages (distributed with LaTeX2e)
% and the \includegraphics macro defined in those packages.
% See the LaTeX Graphics Companion by Michel Goosens, Sebastian Rahtz,
% entire page. There is no need to do explicit centering.

%\begin{figure}
%\includegraphics{fig1.jpg}

%\caption{\label{1} }
% \end{figure}

\begin{acknowledgments}
We acknowledge support of Italian minister of research. We thank P. Ghose, A.S.Majumdar and G. Introzzi for useful discussions.  We thank R. Steni for the realisation of the 
double slit. 
\end{acknowledgments}


\newpage
\section{Figures Captions}
 
Fig.1 The experimental apparatus. A pump laser at 351 nm generates parametric 
down conversion of type I in a lithium-iodate crystal. Conjugated photons at 702 
nm are sent to a double-slit by a system of two piano-convex lenses in a way 
that each photon of the pair crosses a well defined slit. A first photodetector 
is placed at  1.21 m  a second one at 1.5 m from the slit. Both the single 
photon detectors (D) are preceded by an interferential filter at 702 nm (IF) and 
a lens (L) of 6 mm diameter and 25.4 mm focal length. Signals from detectors are 
sent to a Time Amplitude Converter and then to the acquisition system (multi-
channel analyser and counters). 

Fig.2  Coincidences data in the region of interest compared with quantum mechanics predictions.
 On the x-axis we report the position of the first detector respect to the median symmetry axis of the double slit.
 The second detector is kept fixed at -0.055 m. 
The x errors bars represent the width of the lens before the detector. 
A correction for laser power  fluctuations is included.
  
Fig.3 Observed coincidence peak (output of the multi-channel analyser) when 
the centre of the lens is placed 1.7 cm after the median symmetry axis of the 
double slit in the same semiplane of the other photodetector, which is kept at 5.5 cm after the median symmetry axis. 
Acquisition time lasts  17 hours. No background subtraction is done. 
A coincidence peak is clearly visible for a delay between start (first photodector) and stop (second photodector) of 9ns (the delay inserted on the second line signal). 
This result is at variance with the prediction of de Broglie - Bohm theory \cite{4,5,6}.

\end{document}